\documentclass[twocolumn,showpacs,preprintnumbers,amsmath,amssymb,
superscriptaddress]{revtex4-1}

\usepackage{graphicx}
\usepackage{dcolumn}
\usepackage{bm}

\begin{document}


\title{Search for $\alpha$-cluster states in even-even Cr isotopes}

\author{M.~A.~Souza}
\email{marsouza@if.usp.br} 
\affiliation{Instituto de F\'{\i}sica, Universidade de S\~{a}o Paulo, Rua do
Mat\~{a}o, 1371, CEP 05508-090, Cidade Universit\'{a}ria, S\~{a}o
Paulo - SP, Brazil}
\affiliation{Instituto Federal de Educa\c{c}\~{a}o, Ci\^{e}ncia e Tecnologia de
S\~{a}o Paulo - Campus S\~{a}o Paulo, Rua Pedro Vicente, 625, CEP 01109-010,
Canind\'{e}, S\~{a}o Paulo - SP, Brazil}
\author{H.~Miyake}
\email{miyake@if.usp.br}
\affiliation{Instituto de F\'{\i}sica, Universidade de S\~{a}o Paulo, Rua do
Mat\~{a}o, 1371, CEP 05508-090, Cidade Universit\'{a}ria, S\~{a}o
Paulo - SP, Brazil}

\begin{abstract}
The $\alpha + \mathrm{core}$ structure is investigated in even-even Cr isotopes from
the viewpoint of the local potential model. The comparison of $Q_{\alpha}/A$ values
for even-even Cr isotopes and even-even \mbox{$A = 46,54,56,58$} isobars indicates that
$^{46}$Cr and $^{54}$Cr are the most favorable even-even Cr isotopes for \mbox{$\alpha$-clustering}.
The ground state bands of the two Cr isotopes are calculated through
a local $\alpha + \mathrm{core}$ potential with two variable parameters.
The calculated spectra give a very good description of most experimental
$^{46}$Cr and $^{54}$Cr levels. The reduced $\alpha $-widths, rms intercluster separations
and $B(E2)$ transition rates are determined for the ground state bands.
The calculations reproduce the order of magnitude of the available experimental $B(E2)$
values without using effective charges and indicate that the first members of the ground
state bands present a stronger $\alpha$-cluster character. The volume integral per nucleon
pair and rms radius obtained for the $\alpha+^{50}$Ti potential are consistent with those
reported previously in the analysis of $\alpha$ elastic scattering on $^{50}$Ti.
\end{abstract}

                            
                              
\maketitle

\section{Introduction}
\label{Sec:Introduction}

The $\alpha$-cluster structure is present in nuclei of different mass regions and is matter of discussions
with many different approaches. A comprehensive review of this research theme, as well as the nuclear
clustering in general, is found in Ref.~\cite{HIK2012}, and recent advances are described in
Refs.~\cite{HIK2012,B2014}. The $\alpha$-cluster model has been successful in describing
energy levels, electromagnetic transition strengths, $\alpha$-decay widths and $\alpha$-particle elastic
scattering data. Previous studies \cite{BMP2000,BMP1999} have questioned whether the most likely
\mbox{cluster + core} structures are determined strictly by the doubly closed shell for cluster and core.
Our recent article \cite{SM2015} shows that the local potential model (LPM) with
the $\alpha + \mathrm{core}$ interpretation can be applied systematically in intermediate mass nuclei around
the double shell closure at $^{90}$Zr, obtaining a good description of the energy levels and orders of magnitude
of the $B(E2)$ transition rates without the use of effective charges. Additionally, Ref.~\cite{SM2015} shows that
the $\alpha + \mathrm{core}$ potential employed in $^{94}$Mo and $^{96}$Ru are similar to the real parts of the
optical potentials used in other studies for analysis of $\alpha + ^{90}$Zr and $\alpha + ^{92}$Mo elastic
scattering, respectively.

In the $fp$-shell region, $^{44}$Ti is one of the most studied nuclei with the $\alpha + \mathrm{core}$
interpretation, since it has the configuration of $\alpha$-particle plus $^{40}$Ca doubly magic core. Works
on the $\alpha + ^{40}$Ca structure in $^{44}$Ti (examples in Refs.~\cite{MRO88,BMP95,MPH1989,OHS1998,MOR1998})
show good descriptions of the ground state band and $B(E2)$ transition rates.
Neighboring nuclei as $^{43}$Sc, $^{43}$Ti, and others at the beginning of the $fp$-shell have been analyzed
with the LPM \cite{M1984,M1987}, however, considering the effect of noncentral forces that arise with non-zero spin
cores. The $\alpha + \mathrm{core}$ structure in the $fp$-shell region has also been examined in nuclei as $^{42}$Ca
and $^{43}$Sc from the viewpoint of the orthogonality condition model \cite{SO1995,SO1998} and
deformed-basis antisymmetrized molecular dynamics \cite{T2014} with favorable results.

The motivation of this work is the investigation of the $\alpha + \mathrm{core}$ structure in nuclei of the Cr
isotopic chain, which is near the well-studied $^{44}$Ti and presents even-even nuclei without the
\mbox{$\alpha$ + \{doubly closed shell core\}} configuration. The structure of the Cr isotopes has
been analyzed in previous studies with different approaches, such as the shell-model and its variations
\cite{KSH2008,TSU2015,KL2014}, Hartree-Fock or Hartree-Fock-Bogoliubov models \cite{SHY2012,OM2008,SG1981,BKS1973},
and proton-neutron interacting boson model \cite{KL2014}. Also, the $^{48}$Cr nucleus has been described
by the $^{40}\mathrm{Ca} + \alpha + \alpha$ structure with the use of the generator coordinate method \cite{D2002}
and orthogonality condition model \cite{SO2002}. Recently, T.~Togashi {\it et al.}~\cite{TSU2015} investigated
natural and unnatural-parity states in neutron-rich Cr and Fe isotopes using large-scale shell-model calculations,
obtaining good agreement with experimental energy levels. K.~Kaneko {\it et al.}~\cite{KSH2008}
analyzed neutron-rich Cr nuclei using the spherical shell model; the results give a good general
description of the experimental energy levels, and the experimental $B(E2; 2^{+}_1 \rightarrow 0^{+}_1)$
values for $^{52}$Cr and $^{54}$Cr are well reproduced by using effective charges $e_\pi = 1.5 \, e$
and $e_\nu = 0.5 \, e$.

The nuclei of the Cr region have been intensively studied in recent years, since changes in the nuclear
shell structure have been observed with the increase of neutron number in the isotopic chain, raising a
question about the persistence of the traditional magic numbers in neutron-rich nuclei of the $fp$-shell.
A subshell closure at $N = 32$ is indicated by new measurements of nuclear masses, high $2^{+}_1$ energy levels,
and low $B(E2; 0^{+}_1 \rightarrow 2^{+}_1)$ values in nuclei as $^{52}$Ca \cite{WBB2013,GJB2006}, $^{54}$Ti
\cite{JFM2002,DJG2005}, and $^{56}$Cr \cite{BSG2005,PMB2001} compared with neighboring isotopes, and a
subshell closure at $N = 34$ has been observed in $^{54}$Ca \cite{STA2013}. Different studies
\cite{KL2014,JPX2014,CCF2013,KSH2008} aim for an increase of collectivity for neutron-rich nuclei in the
$fp$-shell region near $N = 40$. In this context, the present work contributes to further discussions on
the nuclear structure of this region.

The next section describes a criterion for selection of the preferential nuclei for $\alpha$-clustering in
the set of even-even Cr isotopes. In Section \ref{Sec:Model}, the $\alpha$-cluster model and
$\alpha + \mathrm{core}$ potential are described in detail. In Section \ref{Sec:Results}, an analysis of the
selected nuclei ($^{46}$Cr and $^{54}$Cr) is presented from the standpoint of the LPM. Conclusions are shown
in Section \ref{Sec:Conclusions}.

\section{Selection of preferential nuclei for $\alpha$-clustering}

A preliminary question is the choice of a criterion to determine the most
favorable nuclei for $\alpha$-clustering in a specified set of nuclei.
We use the same criterion employed in our previous work \cite{SM2015} which is based
only on experimental data of binding energy \cite{AWW2012,WAW2012}. An
appropriate quantity for comparing different nuclei is the variation of
average binding energy per nucleon of the system due to the
$\alpha + \mathrm{core}$ decomposition. This value is given by

\begin{equation}
\frac {Q_\alpha}{A_T}=\frac{B_\alpha +B_{\mathrm{core}}-B_T}{A_T}\;, 
\end{equation}

\noindent where $Q_\alpha$ is the $Q$-value for $\alpha$-separation, $A_T$ is
the mass number of the total nucleus and $B_\alpha $, $B_{\mathrm{core}}$ and
$B_T$ are the experimental binding energies of the $\alpha $-cluster, the core
and the total nucleus, respectively. Thus, an absolute (or local) maximum of
$Q_{\alpha}/A_T$ indicates the preferred nucleus for $\alpha $-clustering in
comparison with the rest of (or neighbouring) nuclei in the set.

\begin{figure}
\includegraphics[scale=0.34]{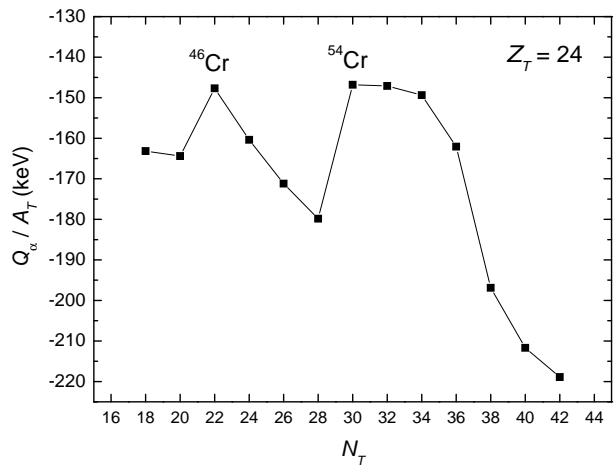}
\caption{$Q_{\alpha}/A_T$ values obtained for the $\alpha $ + core decomposition of even-even
Cr isotopes as a function of the total neutron number $N_{T}$. The $Q_{\alpha}/A_T$ peaks
corresponding to $^{46}$Cr and $^{54}$Cr are indicated.}
\label{Figure_Q_alpha_isotopes}
\end{figure}

This study focuses mainly on the comparison of Cr isotopes. FIG.~\ref{Figure_Q_alpha_isotopes} shows
graphically the values of $Q_{\alpha}/A_T$ for even-even Cr isotopes, where there are two $Q_{\alpha}/A_T$
peaks for $^{46}$Cr ($\approx -147.7$ keV) and $^{54}$Cr ($\approx -146.8$ keV). These nuclei correspond to
$^{42}$Ti and $^{50}$Ti cores with magic numbers of neutrons $N = 20$ and $N = 28$, respectively.
However, the $^{56}$Cr and $^{58}$Cr isotopes have $Q_{\alpha}/A_T$ values ($\approx -147.1$ keV and $-149.4$ keV,
respectively) very close to those of $^{46}$Cr and $^{54}$Cr. This feature suggests the influence of a neutron
subshell closure above $N = 28$ for the core. In the case of $^{58}$Cr, there is a $^{54}$Ti core with
$N_{\mathrm{core}} = 32$, which is related to a subshell closure as discussed in previous studies (see Section
\ref{Sec:Introduction}).

\begin{figure}
\includegraphics[scale=1.1]{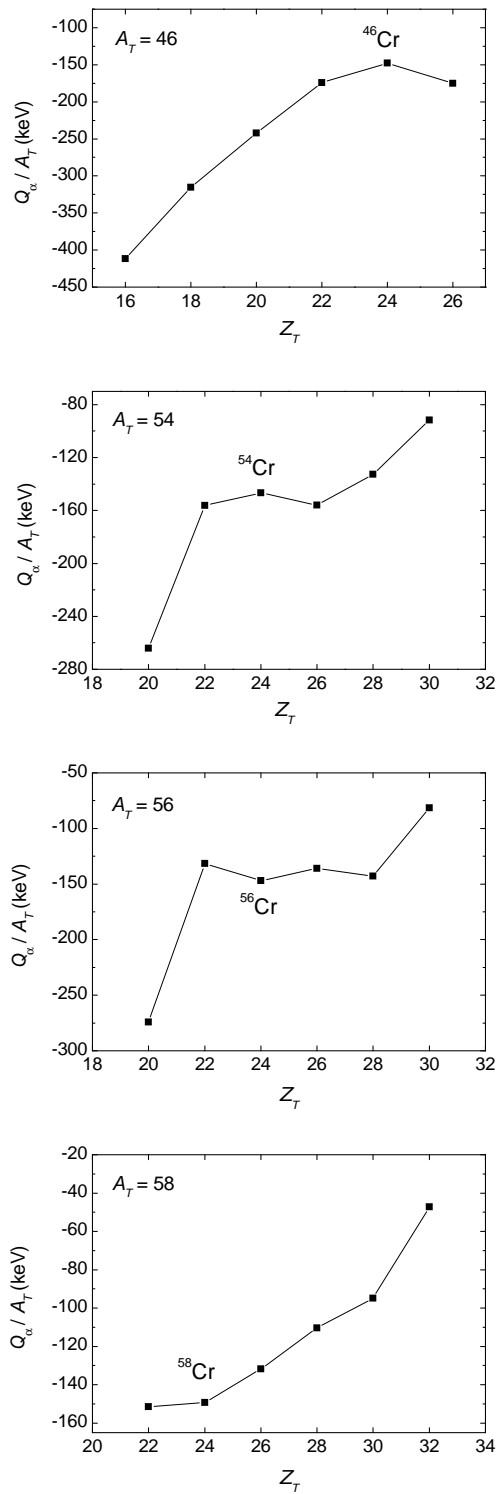}
\caption{$Q_{\alpha}/A_T$ values obtained for the $\alpha $ + core decomposition of even-even
$A=46,54,56,58$ isobars as a function of the total charge number $Z_{T}$. The $Q_{\alpha}/A_T$
values corresponding to $^{46}$Cr, $^{54}$Cr, $^{56}$Cr, and $^{58}$Cr are indicated.}
\label{Figure_Q_alpha_isobars}
\end{figure}

FIG.~\ref{Figure_Q_alpha_isobars} shows graphically the values of $Q_{\alpha}/A_T$ for even-even 
$A = 46,54,56,58$ isobars. An absolute $Q_{\alpha}/A_T$ peak is seen to $^{46}$Cr in the $A = 46$ graph, and
a local peak is seen for $^{54}$Cr in the $A = 54$ graph. However, the $^{56}$Cr and $^{58}$Cr nuclei do not
correspond to absolute or local maxima of $Q_{\alpha}/A_T$ in the $A = 56$ and $A = 58$ graphs. Therefore,
an overall evaluation of the $Q_{\alpha}/A_T$ values in FIGS.~\ref{Figure_Q_alpha_isotopes} and \ref{Figure_Q_alpha_isobars}
implies that $^{46}$Cr and $^{54}$Cr are preferential nuclei for $\alpha$-clustering if they are compared with
other even-even Cr isotopes and respective even-even isobars simultaneously. These two Cr isotopes are
then selected for a more detailed analysis in next sections.

The existence of two $Q_{\alpha}/A_T$ peaks with very close values in FIG.~\ref{Figure_Q_alpha_isotopes}
indicates a transition from $N_{\mathrm{core}} = 20$ to $N_{\mathrm{core}} = 28$ as two preferential numbers
for $\alpha$-clustering in this mass region. However, the variation of $Q_{\alpha}/A_T$ is influenced by the
changes in the nuclear shell structure for neutron-rich nuclei, and also the liquid drop behavior of the
binding energy which is significant for many nuclei. For this reason, in other isotopic chains, the
preferencial number of neutrons of the core may vary in relation to the traditional magic numbers.

\section{$\alpha $-cluster model}
\label{Sec:Model}

The properties of the nucleus are viewed in terms of a preformed
$\alpha $-particle orbiting an inert core. Internal excitations of the
$\alpha$-cluster and the core are not considered in the calculations. The
$\alpha + \mathrm{core}$ interaction is described through a local
phenomenological potential \mbox{$V(r)=V_C(r)+V_N(r)$} containing the Coulomb
and nuclear terms. For the nuclear potential, we adopt the form

\begin{eqnarray}
V_N(r) = -V_{0}\left[1+\lambda\exp\left(-\frac{r^{2}}{\sigma^{2}}\right)\right]
 \left\{ \frac b{1+\exp [(r-R)/a]} \right. {}
                                                          \nonumber\\[4pt]
 {} \left. + \frac{1-b}{\{1+\exp[(r-R)/3a]\}^3}\right\} \;, \qquad \qquad
\label{eq:Nuc_Pot}
\end{eqnarray}

\noindent where $R$ and $\sigma$ are free parameters and $V_0$, $\lambda$, $a$
and $b$ are fixed parameters. The Coulomb potential $V_C(r)$ is taken to be
that of an uniform spherical charge distribution of radius $R_C=R$. The inclusion
of the centrifugal term results in the effective potential

\begin{equation}
V_{\mathrm{eff}}(r)=V(r)+\frac{L\left(L+1\right)\hbar^{2}}{2\mu r^{2}}\;,
\label{eq:Veff}
\end{equation}

\noindent where $\mu$ is the reduced mass of the $\alpha + \mathrm{core}$
system.

The shape employed in eq.~\eqref{eq:Nuc_Pot} is a variation of the modified Woods-Saxon potential W.S.$+$W.S.$^{3}$.
The factor of type \mbox{(1 + Gaussian)} allows the correct reproduction of the 0$^{+}$ bandhead, which is described
roughly with the original W.S.$+$W.S.$^{3}$ potential in previous calculations for other nuclei \cite{BMP95,SM2015}.
The effect of the (1 + Gaussian) factor in the effective potential with $L > 0$ is very weak; therefore, only the
0$^{+}$ level is changed significantly in comparison with the spectrum produced by the simple W.S.$+$W.S.$^{3}$ potential.

The ground state bands of $^{46}$Cr and $^{54}$Cr are calculated with
the fixed values $V_0 = 220$ MeV, $a = 0.65$ fm, $b = 0.3$ and $\lambda = 0.14$,
while $R$ and $\sigma$ are adjusted separately for each nucleus.
The values of $V_0$, $a$ and $b$ are the same used in Refs.~\cite{BMP95,SM2015} to describe the ground state bands of nuclei
of different mass regions with the W.S.$+$W.S.$^{3}$ nuclear potential. Firstly, the parameter $\lambda$ is fitted
to reproduce the 0$^{+}$ bandheads of the ground state bands of $^{20}$Ne, $^{44}$Ti, $^{94}$Mo and $^{212}$Po,
using the corresponding $R$ values obtained from Ref.~\cite{SM2015}. Then the parameters $\sigma$ and $R$ are fitted
to reproduce the experimental 0$^{+}$ and 4$^{+}$ members of the ground state bands of $^{46}$Cr and $^{54}$Cr.
(see TABLE \ref{Tab_Radii}).

\begin{table}
\caption{Values of the parameters $R$ and $\sigma$ for $^{46}$Cr and $^{54}$Cr.}
\label{Tab_Radii}
\begin{ruledtabular}
\begin{tabular}{cccc}
Nucleus & System & $R$ (fm) & $\sigma$ (fm)  \\[2pt] \hline
&  &  \\[-6pt] 
$^{46}$Cr & $\alpha+{}^{42}$Ti & 4.658 & 0.248 \\
$^{54}$Cr & $\alpha+{}^{50}$Ti & 4.674 & 0.210 \\
\end{tabular}
\end{ruledtabular}
\end{table}

The Pauli principle requirements for the $\alpha$ valence nucleons are
introduced through the quantum number

\begin{equation}
G = 2N + L \;,
\end{equation}

\noindent where $N$ is the number of internal nodes in
the radial wave function and $L$ is the orbital angular momentum.
The global quantum number $G$ identifies the bands of states. In this way, the
restriction $G\geq G_{\mathrm{g.s.}}$ is applied, where $G_{\mathrm{g.s.}}$
corresponds to the ground state band. The value $G_{\mathrm{g.s.}}=12$ is
employed for $^{46}$Cr and $^{54}$Cr. This value is obtained from the
Wildermuth condition \cite{WT1977} considering the $(fp)^4$
configuration.

The energy levels and associated radial wave functions are calculated
by solving the Schr\"{o}dinger radial equation for the reduced mass of the
$\alpha + \mathrm{core}$ system.

\section{Results}
\label{Sec:Results}

\begin{figure}
\includegraphics[scale=0.8]{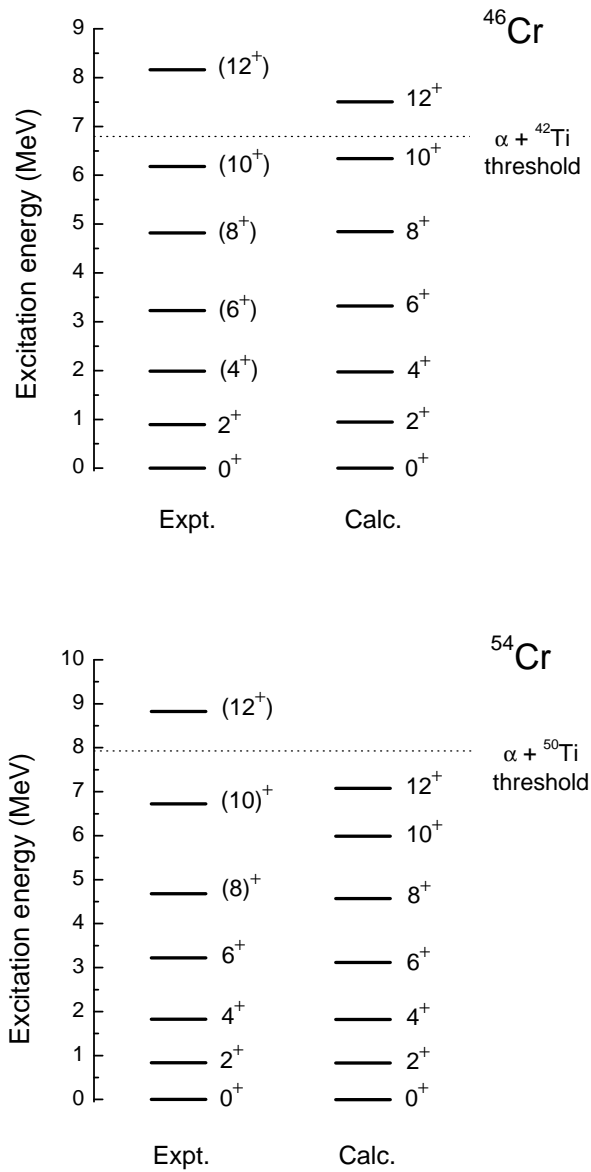}
\caption{Calculated $\alpha + \textrm{core}$ energies for the ground state
bands of $^{46}$Cr and $^{54}$Cr in comparison with experimental
excitation energies. The experimental data are from
Refs.~\cite{S2016,W2000,DJ2014}.}
\label{Figure_G_S_Bands}
\end{figure}

Using the $\alpha + \mathrm{core}$ potential described in Section \ref{Sec:Model},
we have calculated the ground state bands for $^{46}$Cr and $^{54}$Cr.
The results are compared with the corresponding experimental energies in FIG.~\ref{Figure_G_S_Bands}.
The theoretical bands give a very good description of the experimental levels of $^{46}$Cr 
from 0$^{+}$ to 10$^{+}$ and $^{54}$Cr from 0$^{+}$ to 8$^{+}$ (uncertain assignments are indicated in
FIG.~\ref{Figure_G_S_Bands}), and a reasonable description of the higher spin levels, if we consider that
the fixed parameters $V_0$, $a$, $b$ and $\lambda$ have been adjusted to reproduce
the spectra of nuclei of different mass regions. It is gratifying that the mentioned results
are obtained without a dependence on quantum number $L$ in the $\alpha + \mathrm{core}$ potential.

According to Refs.~\cite{DJ2014,FLL1971,FBL1977}, the experimental 0$^{+}$, 2$^{+}$ and 4$^{+}$ levels of
the $^{54}$Cr g.s.~band are populated in the $^{50}$Ti($^{6}$Li,\textit{d})$^{54}$Cr and
$^{50}$Ti($^{16}$O,$^{12}$C)$^{54}$Cr reactions. Such $\alpha$-transfer information reinforce the choice of these
states for comparison with the calculated band. However, there is no mention that the other band members are populated
in the same reactions. New $\alpha$-transfer experiments may be useful to confirm the spins and parities of the levels
above 6$^{+}$ and verify if the levels above 4$^{+}$ are populated by these processes. It is important to observe that
there are several experimental $^{54}$Cr levels above $E_x \approx 4.5$ MeV which are populated in the
$^{50}$Ti($^{16}$O,$^{12}$C)$^{54}$Cr reaction and do not have definite assignments.

It is interesting to compare the $\alpha + \mathrm{core}$ potential of this work with optical potentials
used to describe $\alpha$ elastic scattering. The comparison may be done through the volume integral
per nucleon pair

\begin{equation}
J_{R}=\frac{4\pi}{A_{\alpha} \, A_{\mathrm{core}}}
\int_{0}^{\infty}\, V_{N}(r)\,r^{2}\, dr \;,\label{eq:int_vol}
\end{equation}

\noindent and the root-mean-square (rms) radius associated with the potencial

\begin{equation}
r_{\mathrm{rms},R} = \left[
\frac{\displaystyle \int_{0}^{\infty}\, V_{N}(r)\,r^{4}\, dr}
{\displaystyle \int_{0}^{\infty}\, V_{N}(r)\,r^{2}\, dr} \right]^{1/2}\;.
\label{eq:rms_pot}
\end{equation}

\noindent Eqs.~\eqref{eq:int_vol} and \eqref{eq:rms_pot} refer specifically to the nuclear
real part of the optical potential. TABLE \ref{Tab_Int_vol} shows the $J_R$ and
$r_{\mathrm{rms},R}$ values calculated for the nuclear $\alpha + \textrm{core}$ potentials
of $^{46}$Cr and $^{54}$Cr. As usual, the negative sign of $J_R$ is omitted.

\begin{table}
\caption{Calculated values of the volume integral per nucleon pair
($J_R$) and root-mean-square radius ($r_{\mathrm{rms},R}$) for the
$\alpha + \textrm{core}$ nuclear potentials employed for $^{46}$Cr and $^{54}$Cr.}
\label{Tab_Int_vol}
\begin{ruledtabular}
\begin{tabular}{cccc}
Nucleus & System & $J_R$ (MeV fm$^3$) & $r_{\mathrm{rms},R}$ (fm)\\[2pt] \hline
&  &  &  \\[-6pt] 
$^{46}$Cr & $\alpha+{}^{42}$Ti & 355.0 & 4.314 \\
$^{54}$Cr & $\alpha+{}^{50}$Ti & 301.1 & 4.323 \\
\end{tabular}
\end{ruledtabular}
\end{table}

Ref.~\cite{KBH1982} analyzes the $\alpha$-particle elastic scattering at 140 MeV on $^{50}$Ti
using a double folding nuclear potential with DDM3Y interaction and three different shapes
for the imaginary part of the optical potential; for this case, the values $J_R = 286\textrm{-}290$
\mbox{MeV fm$^3$} and $r_{\mathrm{rms},R} = 4.482$ fm have been obtained. The $J_R$ and
$r_{\mathrm{rms},R}$ values obtained in the present work for $^{54}$Cr are close to the ones
mentioned previously. This comparison shows there is no discrepancy between the $\alpha + \textrm{core}$
potential applied in this work and the real part of the optical potential of Ref.~\cite{KBH1982}.
The $J_R$ and $r_{\mathrm{rms},R}$ values shown in the TABLE \ref{Tab_Int_vol} are also compatible
with the ranges obtained for the same quantities in the analysis of the $\alpha$ elastic scattering on
$^{40}$Ca at different $E_{\alpha,\mathrm{lab}}$ energies \cite{AMA1996}.

The radial wave functions of the states have been determined
to investigate other properties. A bound state approximation has
been used to determine the radial wave functions of the states above
the $\alpha + \mathrm{core}$ threshold. For these calculations,
the depth $V_0$ is adjusted smoothly for each state in order to reproduce the
experimental excitation energies; however, the relative variation of $V_0$ is
below 0.15 \% for the states from 0$^{+}$ to 8$^{+}$, and above 1 \% only for
the 12$^{+}$ state of $^{54}$Cr.

The root-mean-square (rms) intercluster separation is given by

\begin{equation}
\left\langle R^{2}\right\rangle_{G,J}^{1/2}=
\left[\int_{0}^{\infty}r^{2}\, u_{G,J}^{2}(r)\, dr\right]^{1/2}\;,\label{eq:rms}
\end{equation}

\noindent where $u_{G,J}(r)$ is the normalized radial wave function of a
$|G,J\rangle$ state. The value of $\langle R^2 \rangle ^{1/2}$ is seen to
decrease when one goes from the $0^{+}$ state to the highest spin state of each
band (see TABLE \ref{Table_rms_gamma}). This antistretching effect is found in
nuclei of other mass regions where the $\alpha $-cluster structure is studied,
considering different local potential forms \cite{BDV75,MRO88,O1995,SM2015}.
Such a result shows that the inclusion of the \mbox{(1 + Gaussian)} factor in the nuclear
potential does not significantly change this property in relation to the simple
W.S.$+$W.S.$^{3}$ potential.

\begin{table}
\caption{Calculated values for the rms intercluster separation
($\langle R^2 \rangle ^{1/2}$), the reduced $\alpha $-width
($\gamma _\alpha ^2$) and the dimensionless reduced $\alpha $-width
($\theta _\alpha ^2$) for the members of the ground state bands of $^{46}$Cr
and $^{54}$Cr. The channel radii used for the calculation of $\gamma _\alpha ^2$
and $\theta _\alpha ^2$ are obtained from eq.~\eqref{Eq_ch_radius}.}
\label{Table_rms_gamma}
\begin{ruledtabular}
\begin{tabular}{cccc}
&  &  &  \\[-3pt]
\multicolumn{4}{c}{$^{46}$Cr ($\alpha + ^{42}$Ti system)} \\[2pt] \hline
&  &  &  \\[-8pt]
$J^\pi $ & $\langle R^2 \rangle ^{1/2}$ (fm) & $\gamma _\alpha ^2$ (keV) & $\theta _\alpha ^2$ (10$^{-3}$) \\[2pt] \hline
&  &  &  \\[-6pt]
0$^{+}$ & 4.339 & 2.179 & 6.916 \\ 
2$^{+}$ & 4.341 & 2.213 & 7.024 \\ 
4$^{+}$ & 4.299 & 1.690 & 5.364 \\ 
6$^{+}$ & 4.219 & 0.935 & 2.967 \\ 
8$^{+}$ & 4.121 & 0.374 & 1.189 \\ 
10$^{+}$ & 4.010 & 0.084 & 0.268 \\ 
12$^{+}$ & 3.933 & 0.010 & 0.032 \\[2pt] \hline
&  &  &  \\
\multicolumn{4}{c}{$^{54}$Cr ($\alpha + ^{50}$Ti system)} \\[2pt] \hline
&  &  & \\[-8pt]
$J^\pi $ & $\langle R^2 \rangle ^{1/2}$ (fm) & $\gamma _\alpha ^2$ (keV) & $\theta _\alpha ^2$ (10$^{-3}$) \\[2pt] \hline
&  &  &  \\[-6pt]
0$^{+}$ & 4.290 & 0.631 & 2.184 \\ 
2$^{+}$ & 4.290 & 0.637 & 2.205 \\ 
4$^{+}$ & 4.249 & 0.471 & 1.630 \\ 
6$^{+}$ & 4.181 & 0.268 & 0.927 \\ 
8$^{+}$ & 4.089 & 0.101 & 0.350 \\ 
10$^{+}$ & 4.003 & 0.027 & 0.092 \\ 
12$^{+}$ & 3.933 & 0.003 & 0.011
\end{tabular}
\end{ruledtabular}
\end{table}

The radial wave functions are also used for the calculation of the reduced
$\alpha $-width \cite{AY1974,MSV1970}

\begin{equation}
\gamma _{\alpha}^2=\left( \frac{\hbar^2}{2\mu a_{c}}\right) u^2(a_c)\left[
\int_0^{a_c}|u(r)|^2dr\right] ^{-1}\;,
\end{equation}

\noindent where $\mu $ is the reduced mass of the system, $u(r)$ is the radial
wave function of the state and $a_c$ is the channel radius.
In this work, a procedure that avoids an arbitrary choice of channel radius
is used. The value of $a_c$ is given by the relation

\begin{equation}
a_c = 1.295(A_{\alpha}^{1/3} + A_{\mathrm{core}}^{1/3}) + 0.824 \; \mathrm{(fm)} \;,
\label{Eq_ch_radius}
\end{equation}

\noindent obtained from a linear fit \cite{SM2015} that considers other
channel radii used for different $\alpha + \textrm{core}$
systems in the literature. The dimensionless reduced
$\alpha$-width $\theta _\alpha ^2$ is defined as the ratio of
$\gamma _{\alpha}^2$ to the Wigner limit, that is,

\begin{equation}
\theta _\alpha ^2=\frac{2\mu a_{c}^2}{3\hbar ^2}\gamma _{\alpha}^2\;. 
\end{equation}

\noindent Qualitatively, a large value of $\theta _\alpha ^2$ ($\approx 1$) is
interpreted as an evidence of a high degree of $\alpha$-clustering.

The g.s.~bands of $^{46}$Cr and $^{54}$Cr show a rapid decrease of
$\gamma _\alpha ^2$ with the increasing spin (see TABLE \ref{Table_rms_gamma}).
In agreement with the analysis of the rms intercluster separations, the
behavior of $\gamma _\alpha ^2$ suggests a stronger $\alpha $-cluster
character for the first members of these bands. For the two nuclei, the
dimensionless reduced $\alpha $-widths $\theta _\alpha ^2$ present a small
fraction of the Wigner limit, even for the first members of the band.
This is an expected feature for strongly bound $\alpha + \mathrm{core}$ states,
or states above the $\alpha + \mathrm{core}$ threshold and far below the top of the
Coulomb barrier.

It is noted that the $\theta _\alpha ^2$ values for $^{46}$Cr are $\approx 3 \times$
higher than the respective values for $^{54}$Cr. The energy location of the
$\alpha + \mathrm{core}$ threshold has an important influence on this difference. In
the excitation energy scale, the $\alpha + \mathrm{core}$ threshold for $^{54}$Cr is
$\approx 1.1$ MeV higher than the corresponding threshold for $^{46}$Cr. Thus, the radial
wave functions of the $^{54}$Cr states are less intense in the surface region and, consequently,
these states have a lower degree of $\alpha$-clustering in comparison with $^{46}$Cr.

\begin{table}
\caption{Comparison of the calculated $B(E2)$ transition rates for the ground
state bands of $^{46}$Cr and $^{54}$Cr with the corresponding experimental data
\cite{DJ2014,YMA2005}. The calculated values have been obtained without effective charges.}
\label{Table_BE2}
\begin{ruledtabular}
\begin{tabular}{ccc}
&  &  \\[-3pt]
\multicolumn{3}{c}{$^{46}$Cr ($\alpha + ^{42}$Ti system)} \\[2pt] \hline
&  &  \\[-8pt]
$J^\pi $ & $B(E2;J\rightarrow J-2)$ (W.u.) &
 $B(E2)_{\mathrm{exp.}}$ (W.u.) \\[2pt] \hline
&  &  \\[-5pt]
2$^{+}$ & 9.657 & 19(4)\footnote{Deduced from the experimental $B(E2;0_{\mathrm{g.s.}}^{+} \rightarrow 2_{1}^{+})$
  value obtained by K.~Yamada {\it et al.}~\cite{YMA2005}.} \\
4$^{+}$ & 13.045 & --- \\ 
6$^{+}$ & 12.454 & --- \\ 
8$^{+}$ & 10.281 & --- \\ 
10$^{+}$ & 7.057 & --- \\ 
12$^{+}$ & 3.669 & --- \\[2pt] \hline
&  &  \\
\multicolumn{3}{c}{$^{54}$Cr ($\alpha + ^{50}$Ti system)} \\[2pt] \hline
&  & \\[-8pt]
$J^\pi $ & $B(E2;J\rightarrow J-2)$ (W.u.) &
 $B(E2)_{\mathrm{exp.}}$ (W.u.) \\[2pt] \hline
&  &  \\[-5pt]
2$^{+}$ & 7.456 & 14.4(6) \\
4$^{+}$ & 10.049 & 26(9) \\ 
6$^{+}$ & 9.688 & 18(5) \\ 
8$^{+}$ & 8.004 & 12.8(17) \\ 
10$^{+}$ & 5.729 & --- \\ 
12$^{+}$ & 2.966 & --- \\[1pt]
\end{tabular}
\end{ruledtabular}
\end{table}

The model also allows the calculation of the $B(E2)$ transition rates
between the states of an $\alpha$-cluster band. In the case where the cluster
and the core have zero spins, this quantity is given by

\begin{eqnarray}
\lefteqn{B\left(E2;G,J\rightarrow J-2\right)= {} }
                     \nonumber\\[4pt]
& & {} 
\frac{15}{8\pi}\,\beta_{2}^{2}\,\frac{J\left(J-1\right)}
{\left(2J+1\right)\left(2J-1\right)}\left\langle
 r_{J,J-2}^{2}\right\rangle ^{2}\;,
\label{BE2_a}
\end{eqnarray}

\noindent where

\begin{equation}
\left\langle r_{J,J-2}^{2}\right\rangle =
\int_{0}^{\infty}r^{2}\, u_{G,J}(r)\, u_{G,J-2}(r)dr \;, \label{BE2_b}
\end{equation}

\noindent $\beta_{2}$ is the recoil factor, given by

\begin{equation}
\beta_{2}=\frac{Z_{\alpha}A_{\mathrm{core}}^{2}+Z_{\mathrm{core}}A_{\alpha}^{2}}
{\left(A_{\alpha}+A_{\mathrm{core}}\right)^{2}}\;,
\label{BE2_c}
\end{equation}

\noindent $u_{G,J}(r)$ and $u_{G,J-2}(r)$ are the radial wave functions of the
initial $|G,J\rangle$ state and final $|G,J-2\rangle$ state, respectively.

The calculated $B(E2)$ transition rates for the g.s.~bands of $^{46}$Cr and $^{54}$Cr are presented
in TABLE \ref{Table_BE2}. A comparison of the calculated values and experimental data
shows that the model can provide the correct order of magnitude of the experimental
$B(E2)$ values without the use of effective charges. These results may be considered
satisfactory since, in shell-model calculations for nuclei of this mass region
\cite{KSH2008,KL2014,LDR1995,I1981}, substantial effective charges are necessary to reproduce
the experimental data. Furthermore, it is shown that the calculated $B(E2)$ values for $^{54}$Cr
reproduce nicely the increasing or decreasing trend of the experimental data between the $2^{+} \rightarrow 0^{+}$
and $8^{+} \rightarrow 6^{+}$ transitions.

There are few negative parity levels with definite assignments for $^{46}$Cr and $^{54}$Cr, and
the clear identification of negative parity bands is not possible.
Nevertheless, we have calculated the $3^{-}$ level of the $G = 13$ band for a comparison with
experimental energy levels of the two nuclei, applying the depth $V_0 = 238$ MeV used in the simple
W.S.$+$W.S.$^{3}$ potential for the calculation of the negative parity bands of even-even nuclei
around $^{94}$Mo \cite{SM2015}. The energies $E_{x \, \mathrm{calc}}(3^{-}) = 3.444$ MeV and 3.258 MeV
are obtained for $^{46}$Cr and $^{54}$Cr, respectively, to be compared with the experimental excitation
energies 3.1965 MeV (uncertain assignment) and 4.12705 MeV (definite assignment), respectively.
A consistent analysis of the $\alpha + \textrm{core}$ negative parity bands depends on further
experimental data.

\section{Conclusions}
\label{Sec:Conclusions}

The calculation of $Q_{\alpha}/A_T$ values for even-even Cr isotopes and even-even $A = 46,54,56,58$ isobars
indicates that $^{46}$Cr and $^{54}$Cr are the preferential nuclei for \mbox{$\alpha$-clustering} when compared with their
even-even isotopes and isobars simultaneously. The $\alpha$-cluster model gives a good account of the experimental
ground state bands of these two nuclei through a local $\alpha + \mathrm{core}$ potential with two
variable parameters. The nuclear potential with \mbox{(1 + Gaussian)$\times$(W.S.$+$W.S.$^{3}$)} shape
allows the correct reproduction of the 0$^{+}$ bandheads and, additionally, describes the experimental higher
spin levels of the two nuclei very well using a fixed set of four parameters which was successful in
describing the ground state bands in nuclei of different mass regions.

The calculations of the volume integral per nucleon pair and rms radius show that the values for the $^{54}$Cr
nuclear potential are close to those obtained for the real part of the optical potential for $\alpha$-$^{50}$Ti
elastic scattering at 140 MeV \cite{KBH1982}. As the volume integral may vary with the $\alpha + \mathrm{core}$
scattering energy, this comparison should not be seen as complete; however, it is shown that there is no
discrepancy between the $\alpha$-$^{50}$Ti optical potential and the $\alpha + \mathrm{core}$ potential
of this work.

The $B(E2)$ values obtained for $^{46}$Cr and $^{54}$Cr give the correct order of magnitude of
the available experimental data without effective charges. The calculated intercluster rms radii and
reduced $\alpha$-widths suggest that the $\alpha$-cluster character is stronger for
the first members of the ground state bands of $^{46}$Cr and $^{54}$Cr. Therefore, the use of the
(1 + Gaussian) factor in the $\alpha + \mathrm{core}$ nuclear potential does not change significantly this
feature as compared to other $\alpha + \mathrm{core}$ calculations in different mass regions.

New experimental data, especially from $\alpha$-transfer reactions, will be important to complement the
analysis of this work.

\begin{acknowledgments}

The authors thank the members of the Nuclear Spectroscopy with Light Ions Group of University of S\~{a}o
Paulo for the productive discussions. This work was financially supported by Coordena\c{c}\~{a}o de
Aperfei\c{c}oamento de Pessoal de N\'{i}vel Superior (CAPES).

\end{acknowledgments}

\end{document}